\documentclass[10pt,letterpaper]{article}
\usepackage{arxiv}
\usepackage[T1]{fontenc}
\usepackage{textcomp}
\usepackage{amsmath,amssymb,booktabs,graphicx,algorithm,algpseudocode,array}
\usepackage[numbers,sort&compress]{natbib}
\usepackage[hyphens]{url}
\usepackage{microtype}
\usepackage[font=normalsize,labelfont=bf,labelsep=period]{caption}
\usepackage{xcolor}
\definecolor{linkblue}{rgb}{0.21,0.49,0.74}
\usepackage[breaklinks,colorlinks,allcolors=linkblue]{hyperref}
\renewcommand{\shorttitle}{Geometry-Bridge Flow Matching}
\hypersetup{
  pdftitle={Geometry-Bridge Flow Matching for Blind Rigid-Motion Correction in Head Cone-Beam CT},
  pdfauthor={Sungho Yun and Seungryong Cho}
}
\newcommand{\changed}[1]{{#1}}
\title{Geometry-Bridge Flow Matching for Blind Rigid-Motion Correction in Head Cone-Beam CT}
\author{Sungho Yun and Seungryong Cho\\
{\normalfont Department of Nuclear and Quantum Engineering}\\
{\normalfont Korea Advanced Institute of Science and Technology (KAIST)}\\
{\normalfont Daejeon 34141, South Korea}}
\date{}
\begin{document}
\maketitle
\begin{abstract}
Rigid motion of a patient head during head cone-beam CT (CBCT) scan can degrade image quality and complicate recovery of accurate anatomical structures from the measured projections. We present a 3D flow-matching prior approach that learns the image changes induced by motion. Its geometry-bridge connects motion-corrupted and motion-free Feldkamp--Davis--Kress (FDK) reconstructed images of the same source volume through progressively reduced simulated motion. The forward projector's geometry derivative supplies analytic velocity targets along this path. A plug-and-play loop encompasses the learned image updates together with spline-based pose estimation, conjugate-gradient data consistency, and image total-variation regularization. Both the refined image volume and the FDK reconstructed image volume at the estimated poses are evaluated. On simulated 360-view scans from 30 held-out CQ500 patients with nominal per-component motion amplitudes within 10\,mm\,/\,10\textdegree{}, the framework achieved state-of-the-art image reconstruction quality and motion accuracy in comparison with other methos such as learned autofocus and diffusion-based joint reconstruction. Mean peak signal-to-noise ratio (PSNR) reached 37.1\,dB and mean detector-plane reprojection error (RPE) was 0.27\,mm. With matched endpoints at the inference settings, the geometry bridge improved PSNR by 1.6\,dB and reduced RPE by 33\% relative to image-linear training. With 50 prior updates and coarse-to-fine motion estimation, total processing time averaged 9.1 minutes per patient on an RTX A6000.

\end{abstract}

\begin{figure}[tbp]
\centering
\includegraphics[width=0.75\linewidth]{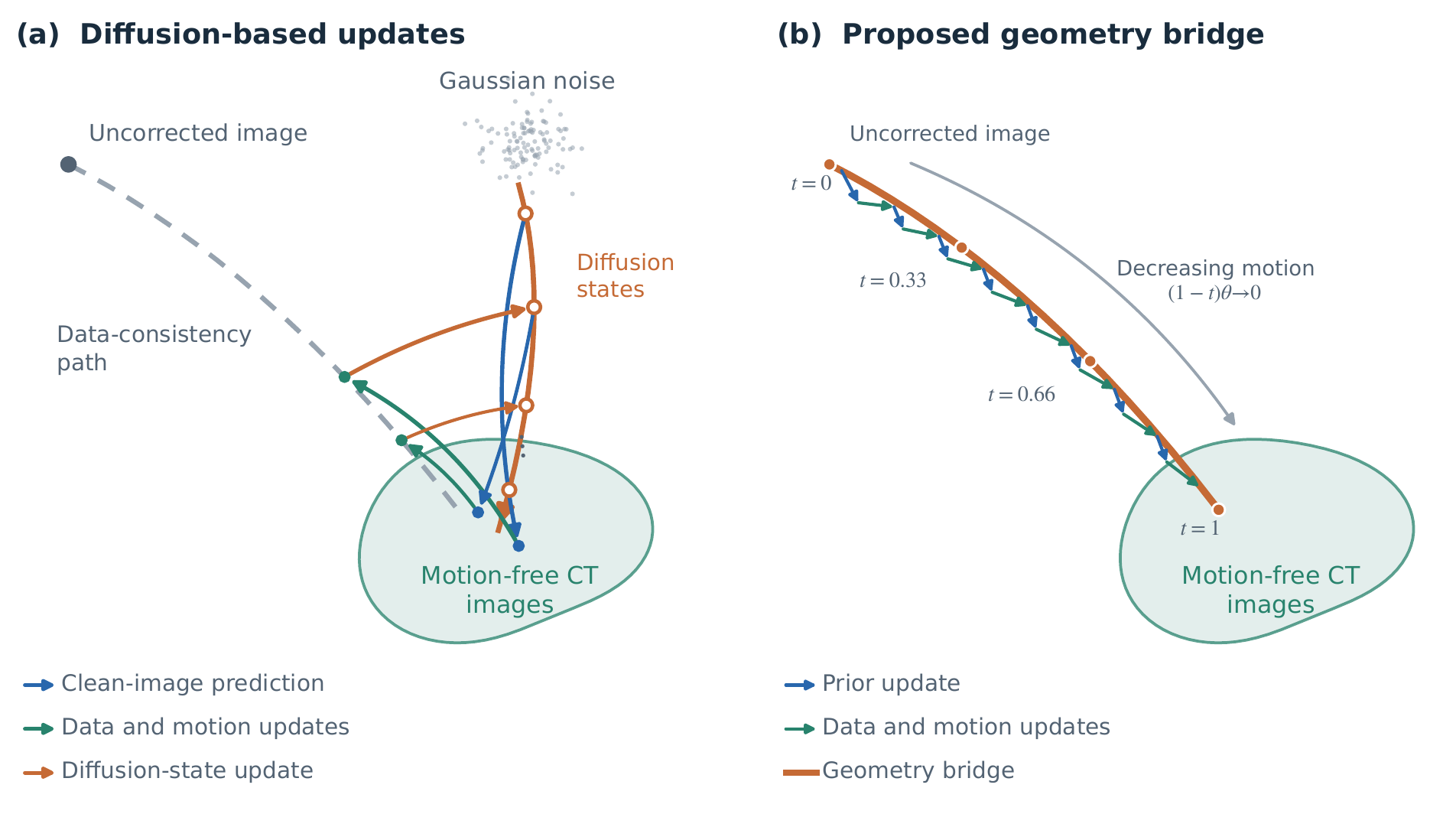}
\caption{Schematic update paths for (a) diffusion-based reconstruction, illustrated by JRM-ADM, and (b) the proposed geometry bridge. Green regions denote motion-free CT images; the proposed training endpoint is motion-free FDK.}
\label{fig:1}
\end{figure}

\section{Introduction}\label{i.-introduction}

Cone-beam CT is widely used for head imaging in dentistry, head-and-neck image-guided radiotherapy, interventional suites and point-of-care settings. Head motion violates the static-object assumption of analytic image reconstruction producing typical image artifacts such as shades, streaks, and double contours. A six-degree-of-freedom (6-DoF) rigid pose per projection view provides a useful model of such skull motion. Recovering these poses from projections alone is challenging because the image itself used for motion estimation is corrupted. Reliable recovery therefore requires image estimates that support pose fitting for image reconstruction.

Image-domain methods directly correct corrupted reconstructions without estimating acquisition poses \cite{ref1}. Autofocus methods instead optimize reconstruction quality, using classical criteria such as sharpness \cite{ref2} or learned estimates of visual information fidelity (VIF) degradation maps \cite{ref3} and scalar VIF scores \cite{ref4}. These methods provide efficient motion estimation, but heavily depend on how reliable the quality of surrogate guide is for optimization particularly when the motion is substantially large altering the reconstructed anatomy.

Another approach jointly reconstructs the image and the estimates of motion by coupling a learned image prior to projection-domain data fidelity. Diffusion-based inverse solvers can provide a framework for combining learned denoising with measurement-driven updates \changed{\cite{ref5,ref6,chung2023diffusionmbir,hyun2026diffusioninr,kawar2022ddrm,wang2023ddnm}}. For sparse-view head CBCT, De Paepe et al. \cite{ref7} introduced JRM-ADM, which combines a wavelet-domain diffusion prior (WDM \cite{ref8}) with joint image and motion updates. Image refinement balances projection fidelity against a quadratic penalty toward the denoised prediction; the refined image then drives pose estimation. However, the prior is learned along a Gaussian denoising path \changed{\cite{ho2020ddpm}}, independently of the image changes induced by motion correction. As pose estimates improve, data fidelity favors consistency with the refined geometry, while prior regularization continues to favor the denoised prediction. Reconstruction therefore remains a compromise between the solution supported by the refined poses and the image favored by the prior.

We address this mismatch with a 3D flow-matching prior \cite{ref9} trained on a prescribed geometry bridge. The bridge connects motion-corrupted and motion-free FDK reconstructions of the same volume by progressively reducing the estimated acquisition motion. With the underlying anatomy held fixed, its intermediate states capture the image changes associated with reducing motion artifacts. The forward projector's geometry derivative provides analytic velocity targets that describe how each intermediate reconstruction changes as acquisition motion is reduced. During inference, the learned velocity acts directly on the current reconstruction, supplying pose estimation with an image update trained to reduce motion artifacts. The refined poses, in turn, allow data-fidelity updates to reconstruct the volume using more accurate acquisition geometry. These learned and measurement-driven updates work cooperatively toward finer motion correction, aligning the prior with pose-dependent data fidelity to address the reconstruction compromise described above (Fig.~\ref{fig:1}).

Our contributions are threefold. First, we formulate a geometry bridge and analytic velocity supervision for learning a motion-free(?) 3D image prior (Sec.~\ref{b.-geometry-bridge-flow-matching}). Second, we integrate this prior with projection-based pose fitting and image refinement in a blind reconstruction loop that recovers both a refined volume and unknown acquisition poses (Sec.~\ref{c.-blind-rigid-motion-correction-loop}). Third, on 30 held-out patients under a common simulated 360-view protocol, we demonstrate state-of-the-art image quality and motion accuracy relative to the evaluated learned autofocus and JRM-ADM methods. FDK reconstructions at the estimated poses complement direct motion metrics by assessing motion compensation without learned image refinement. A matched comparison with an image-linear bridge isolates the effect of the training path and its velocity targets, while an alternative pose-fitting scheme tests whether the gains persist across motion estimators (Secs.~\ref{a.-reconstruction-quality}--\ref{c.-ablation-study}).

Source code is available at \url{https://github.com/dbstjstod1/Flow_matching_motion_3D}.

\begin{figure}[tbp]
\centering
\includegraphics[width=0.75\linewidth]{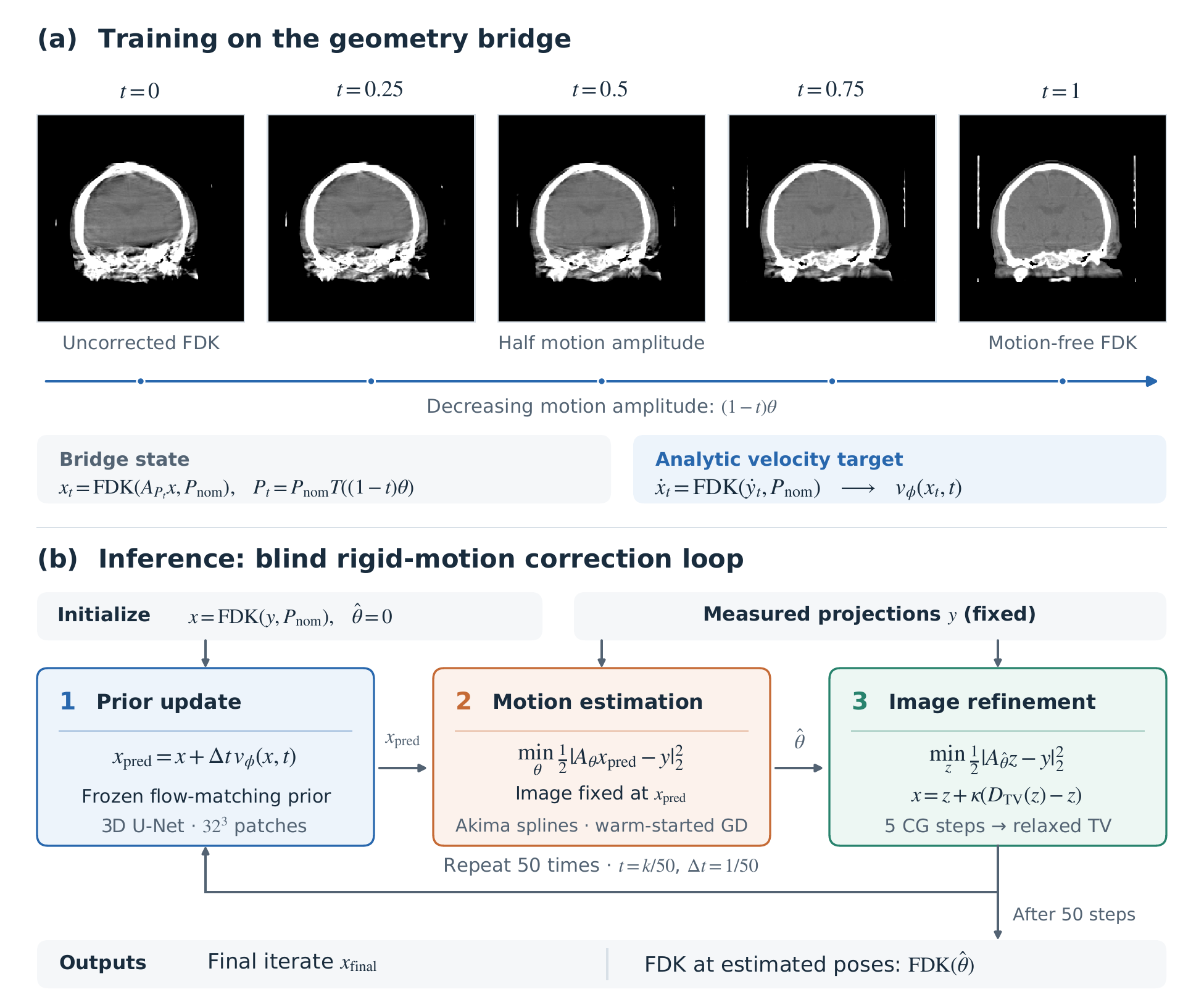}
\caption{Geometry-bridge training and blind inference. (a) FDK reconstructions of the same volume at decreasing motion amplitude \((1 - t)\theta\) provide analytic velocity supervision. (b) Learned image updates alternate with pose fitting and conjugate-gradient (CG)/TV refinement using fixed measurements.}
\label{fig:2}
\end{figure}

\section{Methods}\label{ii.-methods}

\subsection{Forward model and problem formulation}\label{a.-forward-model-and-problem-formulation}

Let \(x\) denote an attenuation volume, or image to reconstruct, and \(y\) the log-transformed projections acquired over \(V\) views on a nominal circular orbit \(P_{\mathrm{nom}}\). The rigid pose at view \(v\) is \(\theta_{v} \in \mathbb{R}^{6}\), comprising a translation vector \((t_{x},t_{y},t_{z})\) and a rotation vector \((r_{x},r_{y},r_{z})\) in axis--angle representation. Translations are in millimeters and rotation vectors in radians, with the rotation center at the volume center. We denote estimated poses by \(\hat{\theta}\). Writing its transform as \(T(\theta_{v})\), the acquisition geometry is \(P(\theta)\lbrack v\rbrack = P_{\mathrm{nom}}\lbrack v\rbrack T(\theta_{v})\) and the forward model is \(y = A_{P(\theta)}x\). We abbreviate this projector as \(A_{\theta}\) and write \(FDK(\theta) = FDK(y,P(\theta))\) for analytic reconstruction at a given trajectory \cite{ref10}.

We recover the volume and acquisition poses by combining projection-domain data fidelity, a learned motion-correction prior and total-variation (TV) regularization. We express this balance as a regularized recovery problem:

\begin{equation}\label{eq:recovery}
\min_{x,\theta}\;\frac12\lVert A_\theta x-y\rVert_2^2+\mathcal R_{\mathrm{TV}}(x)+\mathcal R_{\mathrm{prior}}(x).
\end{equation}

Here \(\mathcal{R}_{\mathrm{TV}}(x) = \sum_{d = 1}^{3} \parallel D_{d}x \parallel_{1}\), where \(D_{d}\) denotes derivatives along spatial axis \(d\). The term \(\mathcal{R}_{\mathrm{prior}}\) represents the implicit regularization supplied by the learned velocity field \(v_{\phi}(x,t)\) (Sec.~\ref{b.-geometry-bridge-flow-matching}); it does not assume an explicit scalar potential. The reconstruction algorithm implements this balance through plug-and-play updates \changed{\cite{ref11,martin2025pnpflow,zhang2017ircnn,zhu2023diffpir}}, defined in Sec.~\ref{c.-blind-rigid-motion-correction-loop}.

\subsection{Geometry-bridge flow matching}\label{b.-geometry-bridge-flow-matching}

\changed{Figure~\ref{fig:2} summarizes the training construction and its use during inference.}

For a training volume \(x\) and simulated motion \(\theta\), the geometry bridge reduces acquisition motion while keeping the FDK reconstruction geometry nominal:

\begin{equation}\label{eq:bridge}
\begin{aligned}
P_t&=P_{\mathrm{nom}}T((1-t)\theta),\qquad y_t=A_{P_t}x,\\
x_t&=\operatorname{FDK}(y_t,P_{\mathrm{nom}}).
\end{aligned}
\end{equation}

Scaling the translation and rotation vectors by \((1 - t)\) reduces both to zero as \(t\) increases from 0 to 1. The initial state is the uncorrected reconstruction, \(x_{0} = FDK(y,P_{\mathrm{nom}})\), for the simulated scan \(y = A_{P(\theta)}x\). The endpoint \(x_{1} = FDK(A_{P_{\mathrm{nom}}}x,P_{\mathrm{nom}})\) is its motion-free FDK counterpart. Each intermediate state is reconstructed from a scan simulated at the corresponding motion amplitude. The anatomy remains fixed, so the bridge describes how motion changes the reconstructed image throughout the path (Fig.~\ref{fig:2}(a)).

Since FDK is linear against its projection input and its reconstruction geometry is fixed, the path tangent is

\begin{equation}\label{eq:tangent}
\begin{aligned}
\dot x_t&=\operatorname{FDK}(\dot y_t,P_{\mathrm{nom}}),\\
\dot y_t&=\left.D_P(A_Px)\right|_{P=P_t}[\dot P_t],\\
\dot P_t&=P_{\mathrm{nom}}\frac{d}{dt}T((1-t)\theta).
\end{aligned}
\end{equation}

The tangent specifies the instantaneous image change induced by reducing simulated motion, providing a velocity target tied to the acquisition model. Here \(D_{P}(A_{P}x)\lbrack{\dot{P}}_{t}\rbrack\) is the directional geometry derivative of the forward projector. The rigid-transform derivative is analytic under this parameterization, and the projector derivative is obtained by differentiating the discrete forward model used to generate the bridge. Each sampled volume and motion trajectory defines an image path \(x_{t}\); their distribution induces the probability path used for flow matching \cite{ref9}. We minimize the mean squared velocity error \(\mathcal{L}_{\mathrm{FM}}(\phi) = \mathbb{E}_{x,\theta,t}\lbrack \parallel v_{\phi}(x_{t},t) - {\dot{x}}_{t} \parallel_{2}^{2}\rbrack\), with \(t \sim \mathcal{U}\lbrack 0,1\rbrack\). The simulated motion is used only to construct training pairs and velocity targets; inference requires neither the true motion nor the motion-free image. A 3D U-Net \changed{\cite{cicek2016unet3d}} parameterizes the flow-matching velocity field \(v_{\phi}(x,t)\).

\subsection{Blind rigid-motion correction loop}\label{c.-blind-rigid-motion-correction-loop}

At inference, the prior learned from these simulated paths is coupled to the measured projections through a plug-and-play loop. Starting from the uncorrected reconstruction and zero motion, each of \(N = 50\) iterations applies a prior update, fits the motion using the updated image, and refines the reconstruction using the measured projections (Algorithm~\ref{alg:loop}). The network remains frozen, and flow time schedules the prior updates; it does not estimate the remaining motion amplitude. Motion fitting and data consistency use the fixed projections \(y\) throughout inference.

\begin{algorithm}[t]
\caption{Blind rigid-motion correction loop with the geometry-bridge prior}
\label{alg:loop}
\normalsize
\textbf{Input:} projections $y$, nominal geometry $P_{\mathrm{nom}}$, frozen prior $v_\phi$, number of steps $N$
\begin{algorithmic}[1]
\State $x\gets\operatorname{FDK}(y,P_{\mathrm{nom}})$, $\hat\theta\gets0$, $\Delta t\gets1/N$
\For{$k=0,\ldots,N-1$}
\State $t\gets k/N$, $x_{\mathrm{pred}}\gets x+\Delta t\,v_\phi(x,t)$
\State $\hat\theta\gets\operatorname{FitMotion}(x_{\mathrm{pred}},y;\hat\theta)$
\Statex \hspace{\algorithmicindent}\textit{(warm-started spline GD)}
\State $z\gets\operatorname{CG}_5(A_{\hat\theta},y;z_0=x_{\mathrm{pred}})$
\Statex \hspace{\algorithmicindent}\textit{(five data-update iterations)}
\State $x\gets z+\kappa\,(D_{\mathrm{TV}}(z)-z)$ \textit{(relaxed TV update)}
\EndFor
\State \Return final iterate $x_{\mathrm{final}}=x$ and $\operatorname{FDK}(\hat\theta)$
\end{algorithmic}
\end{algorithm}


Previous CBCT motion-correction methods model each pose component with a temporal spline and optimize its node values or coefficients \cite{ref2,ref12,ref4}. This represents a continuous motion trajectory with fewer unknowns than independent poses at every projection view. Following this approach, we use an Akima spline \cite{ref13} through 30 uniformly spaced nodes for each of the six motion components and optimize the node values by gradient descent (GD) \cite{ref3}. With the prior-updated image fixed, GD reduces \(\parallel A_{\theta}x_{\mathrm{pred}} - y \parallel_{2}^{2}\) using the projector's geometry gradient. The fitted nodes are retained between flow steps.

At the fitted poses, CG starts from \(x_{\mathrm{pred}}\) and reduces the projection residual, allowing measurement-driven corrections to build on the learned update. The TV denoising operator \(D_{\mathrm{TV}}\) then applies normalized gradient steps to a smooth approximation of \(\mathcal{R}_{\mathrm{TV}}\), with relaxation parameter \(\kappa\) controlling its contribution to the next iterate. The denoising and relaxation settings determine the effective TV strength. The motion and CG subproblems are solved approximately, and the resulting images need not remain on the training bridge. The bridge defines the prior's training path; measured data guide the reconstruction during inference.

The loop returns the final iterate \(x_{\mathrm{final}}\) and \(FDK(\hat{\theta})\), reconstructed from the measured projections at the estimated poses. The latter applies no learned image update, although its poses are estimated with the aid of the prior.

\section{Experimental setup}\label{iii.-experimental-setup}

\subsection{Data and motion simulation}\label{a.-data-and-motion-simulation}

We use 230 patients from the CQ500 head CT collection \cite{ref14}, divided at the patient level into 150 for training, 50 for validation and 30 for testing. Each patient contributes one thin-slice CT series. We simulate 360-view head CBCT scans using the geometry of Thies et al. \cite{ref3} and reconstruct on a \(256^{3}\) grid with 1\,mm voxels. Projections are generated on a finer grid to reduce dependence on the reconstruction discretization, with no added measurement noise.

Motion is simulated with a zero-centered Akima spline through 10 nodes per component. Training covers nominal amplitudes up to 15\,mm for translation and 20\textdegree{} for rotation; testing uses 10\,mm\,/\,10\textdegree{}. Here, amplitude denotes the full width of the control-point sampling interval: test nodes are drawn within \(\pm 5\)\,mm or \(\pm 5^{\circ}\) before zero-centering. The realized peak-to-peak range can differ because of finite sampling and interpolation. Each test patient has one fixed motion trajectory, shared by all methods together with the same projection data, enabling paired comparisons.

\subsection{Network and optimization settings}\label{b.-network-and-optimization-settings}

The 3D U-Net predicts velocities from \(32^{3}\) image patches. A downsampled view of the whole volume and spatial coordinates provide global anatomical context, following \cite{ref15}, while patch-wise processing limits memory use. At each outer step, predictions from two shifted patch tilings are averaged to form the volume velocity, and global context is recomputed from the current reconstruction. The prior is trained for 500,000 iterations using AdamW \changed{\cite{loshchilov2019adamw}} and then kept fixed during reconstruction. Forward projection and backprojection use LEAP \cite{ref16}, with a custom implementation of the forward projector's geometry derivatives for prior training and motion estimation. Inference runs for 50 outer steps, each with 200 inner pose-fitting steps. Each inner step samples 24 views. Pose fitting uses a factor-two downsampled volume and detector for the first 25 outer steps, then full resolution; CG and TV refine the image at full resolution throughout. Pose step sizes are selected by minimizing mean RPE on three preselected validation patients.

\subsection{Evaluation metrics and protocol}\label{c.-evaluation-metrics-and-protocol}

We assess image quality using peak signal-to-noise ratio (PSNR) and the structural similarity index (SSIM) \cite{ref17} against the ground-truth CT, after rigid alignment and within the measured field of view. Both metrics are computed in attenuation units without display-window clipping, using the maximum ground-truth attenuation within the evaluation mask as the PSNR peak and SSIM data range. SSIM uses a \(7 \times 7 \times 7\) uniform window. Motion accuracy is measured by reprojection error (RPE) \cite{ref3}: the mean detector-plane displacement of fixed 3D reference points, averaged over points and views and reported in detector millimeters. We also report the mean absolute error (MAE) of each translation-vector and rotation-vector component.

A common rigid change of the object coordinate frame and acquisition geometry leaves the projections unchanged, so the absolute global pose is not identifiable. To remove this coordinate-frame ambiguity, each trajectory is expressed in a zero-mean pose frame by applying a single rigid transform across all views, chosen so that mean translation and mean rotation vector vanish, without fitting to ground-truth motion. Cohort results are reported as mean \(\pm\) standard deviation.

Both output types use the ground-truth CT as the image-quality reference. FDK at the true poses provides an additional analytic reference using the same motion-corrupted scan with known acquisition geometry.

\subsection{Comparison methods}\label{d.-comparison-methods}

\paragraph{Learned autofocus \cite{ref3}.}\label{learned-autofocus-3.}

This method estimates motion by optimizing a learned image-quality score and returns FDK at the estimated poses. Its network predicts a voxelwise VIF-derived degradation map; its spatial average guides gradient descent on Akima spline node values. The pose objective uses this image-quality score without a projection-domain data-fidelity term. We reimplement the published method and train the quality network on our 150 training patients with the same 15\,mm\,/\,20\textdegree{} amplitude limits as the proposed prior.

\paragraph{JRM-ADM \cite{ref7}.}\label{jrm-adm-7.}

This method jointly reconstructs the image and estimates motion using a wavelet-domain diffusion prior \cite{ref8}. It alternates diffusion-guided image refinement with cubic B-spline motion fitting. Image updates balance projection fidelity against a penalty toward the denoised prediction; the refined image then drives pose estimation. We use the authors' released sampler and optimizers, with the prior retrained on our 150 training patients under their noise schedule. Both comparison methods are evaluated under our common 360-view protocol.

\subsection{Ablation settings}\label{e.-ablation-settings}

\paragraph{Training bridge.}\label{training-bridge.}

To test the role of the training path, we replace the geometry bridge with image-linear interpolation, \(x_{t} = (1 - t)x_{0} + tx_{1}\), supervised by \(x_{1} - x_{0}\) \changed{\cite{liu2023rectifiedflow}}. The endpoints, network, training budget and inference settings are matched. Only the intermediate training states and their velocity targets differ, isolating the choice of training path within the same reconstruction loop.

\paragraph{Motion-estimation scheme.}\label{motion-estimation-scheme.}

To test dependence on the pose-fitting scheme, we replace Akima + GD with 20-control-point cubic B-splines optimized by RMSprop, adapting the pose-fitting scheme of JRM-ADM \cite{ref7}. The geometry-bridge prior and other inference settings are fixed, with step sizes selected by the same validation criterion. This comparison changes the spline representation, control-point count and optimizer together.

\section{Results}\label{iv.-results}

\subsection{Reconstruction quality}\label{a.-reconstruction-quality}

\begin{figure}[tbp]
\centering
\includegraphics[width=1\linewidth]{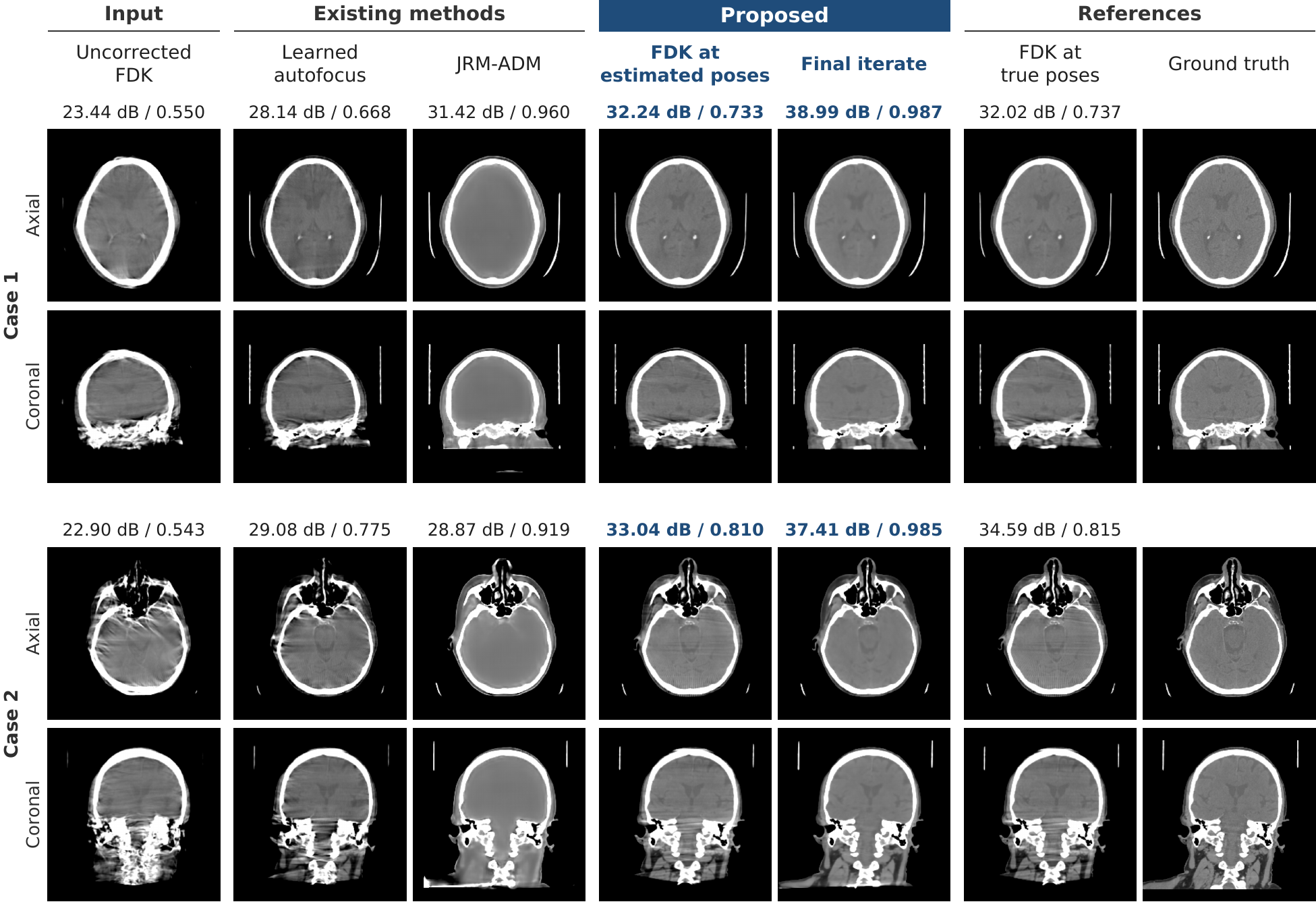}
\caption{Axial and coronal reconstructions for two test patients. FDK uses estimated or true poses of the same motion-corrupted scan. Display window: level 40\,HU, width 420\,HU. Labels give volumetric PSNR (dB) / SSIM after rigid alignment to ground truth within the measured field of view.}
\label{fig:3}
\end{figure}

\begin{figure}[tbp]
\centering
\includegraphics[width=0.8\linewidth]{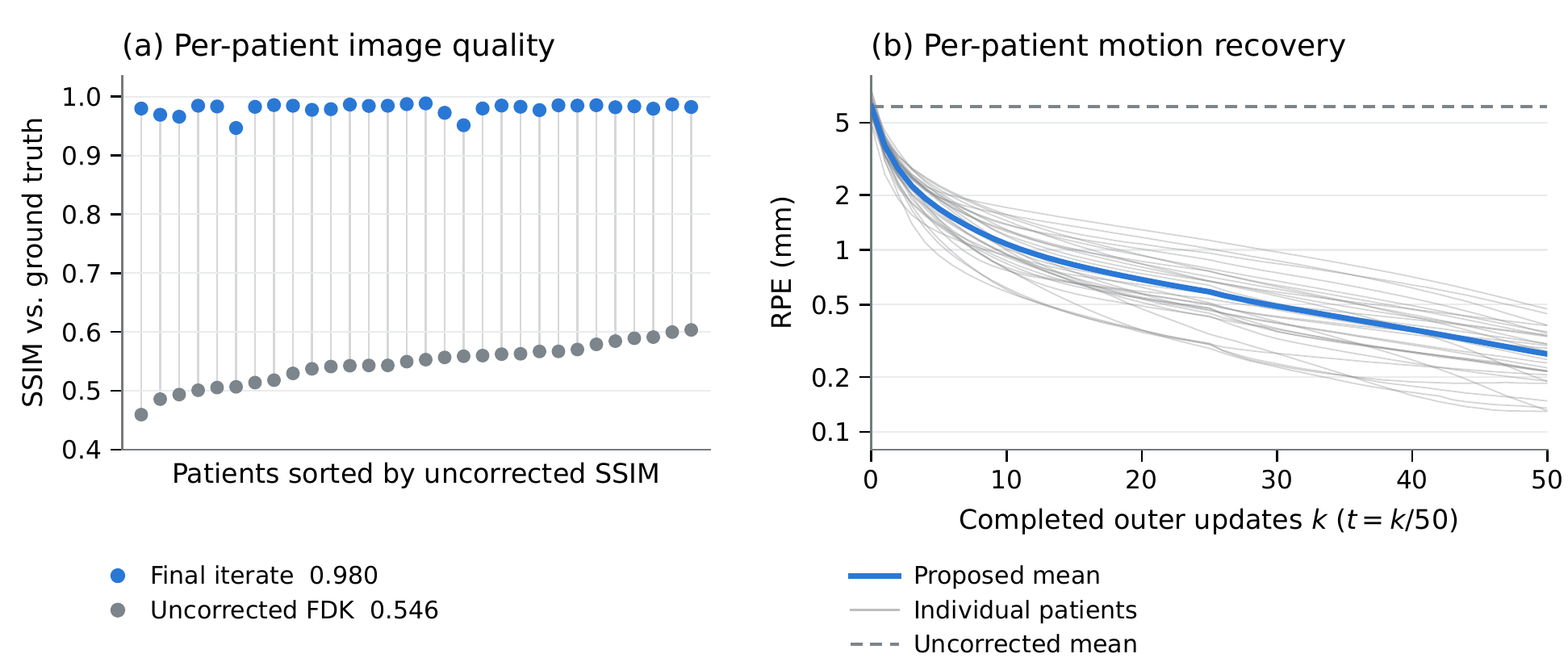}
\caption{Proposed image quality and motion recovery across 30 patients. (a) Paired SSIM scores, ordered by uncorrected SSIM; legend values are cohort means. (b) Individual and mean RPE on a logarithmic scale. Update zero denotes zero-motion initialization.}
\label{fig:4}
\end{figure}

The proposed final iterate achieved the highest mean PSNR and SSIM among the evaluated methods, reaching 37.12\,dB and 0.980, respectively (Table~\ref{tab:1}). Figure~\ref{fig:3} compares reconstructions for two patients, and Fig.~\ref{fig:4} shows cohort image quality and motion recovery.

\begin{table}[tbp]
\caption{Reconstruction quality and motion accuracy (30 patients, 360 views; mean \(\pm\) standard deviation). Each method's FDK and refined outputs share the same estimated poses.}
\label{tab:1}
\centering
\normalsize
\setlength{\tabcolsep}{8pt}
\begin{tabular}{@{}>{\raggedright\arraybackslash}p{125pt}ccc@{}}
\toprule
Method / output & PSNR (dB) & SSIM & RPE (mm) \\
\midrule
Uncorrected FDK & 23.29 \(\pm\) 0.92 & 0.546 \(\pm\) 0.035 & 6.12 \(\pm\) 0.64 \\
Learned autofocus \cite{ref3} & 29.22 \(\pm\) 1.09 & 0.727 \(\pm\) 0.032 & 2.26 \(\pm\) 0.52 \\
JRM-ADM \cite{ref7}, FDK(\(\hat{\theta}\)) & 29.45 \(\pm\) 1.62 & 0.715 \(\pm\) 0.044 & 0.84 \(\pm\) 0.55 \\
JRM-ADM, final output & 29.64 \(\pm\) 2.64 & 0.912 \(\pm\) 0.038 & 0.84 \(\pm\) 0.55 \\
Proposed, FDK(\(\hat{\theta}\)) & 31.93 \(\pm\) 0.89 & 0.753 \(\pm\) 0.032 & 0.27 \(\pm\) 0.09 \\
\textbf{Proposed,} \(x_{\mathrm{final}}\) \textbf{(final iterate)} & \textbf{37.12} \(\pm\) \textbf{1.65} & \textbf{0.980} \(\pm\) \textbf{0.010} & \textbf{0.27} \(\pm\) \textbf{0.09} \\
\bottomrule
\end{tabular}
\end{table}

The gains include both improved motion compensation and additional image refinement. Analytic reconstruction at the proposed poses reached 31.93\,dB PSNR and 0.753 SSIM. It resembles FDK at the true poses in the displayed cases, with residual artifacts in both analytic images. The final iterate improves on \(FDK(\hat{\theta})\) by 5.2\,dB and 0.226 SSIM, showing the benefit of iterative image refinement beyond analytic motion compensation.

FDK at the proposed poses also outperformed analytic reconstructions based on the comparison methods' pose estimates. FDK at the JRM-ADM poses reached 29.45\,dB PSNR and 0.715 SSIM; the proposed FDK gained 2.5\,dB on average. Relative to learned autofocus, its mean gains were 2.7\,dB and 0.026 SSIM.

These improvements were consistent across patients. The proposed final iterate had higher PSNR than JRM-ADM in all 30 patients and exceeded both learned autofocus and uncorrected FDK in PSNR and SSIM in every patient. For analytic outputs, proposed PSNR was higher than FDK at the JRM-ADM poses in 28 patients; compared with autofocus, PSNR was higher in all 30 and SSIM in 24. Qualitatively, JRM-ADM suppressed much of the motion artifact but lost soft-tissue contrast, while autofocus retained streaks, particularly in the coronal slices (Fig.~\ref{fig:3}).

\subsection{Motion recovery}\label{b.-motion-recovery}

\begin{figure}[tbp]
\centering
\includegraphics[width=0.8\linewidth]{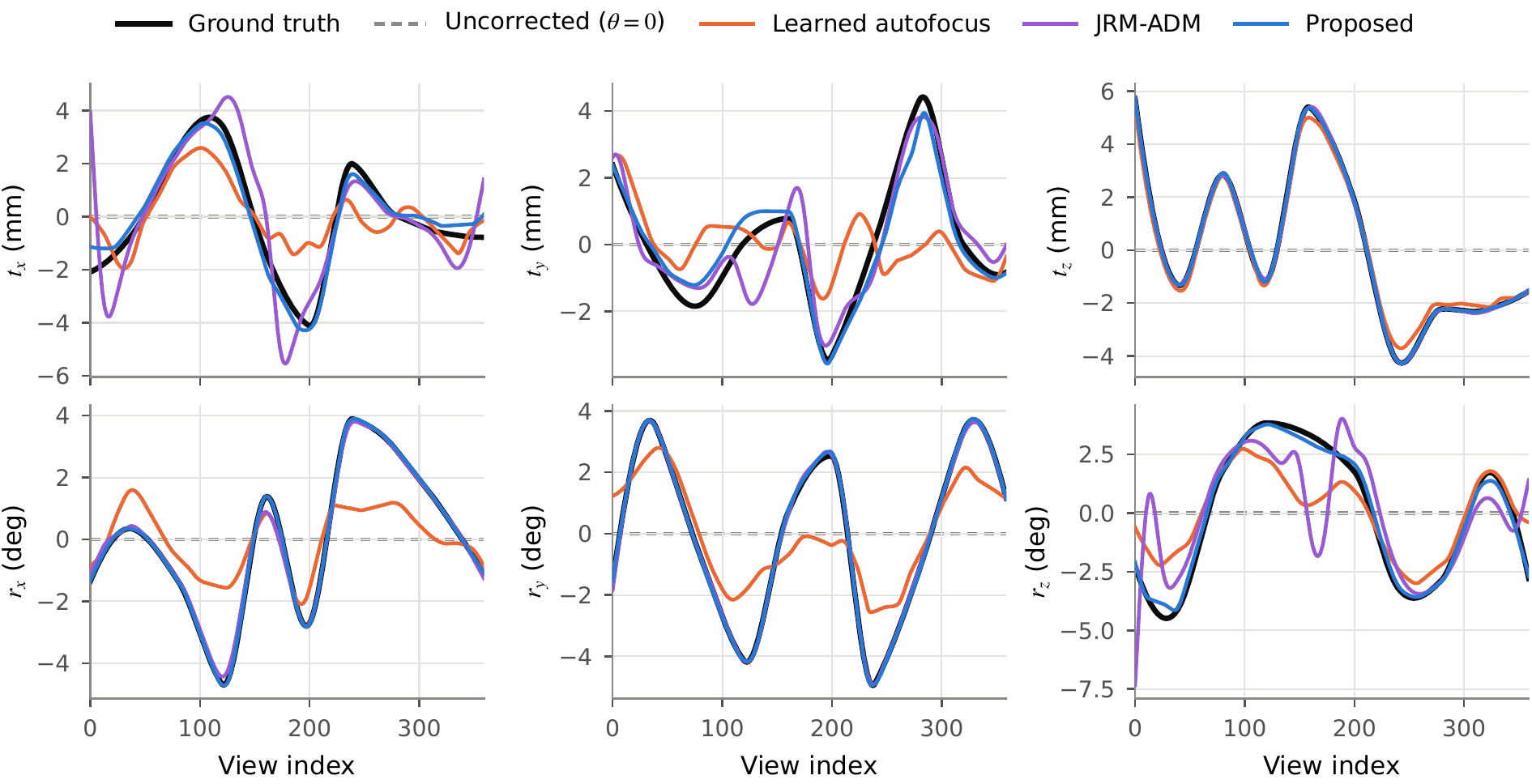}
\caption{Six-component motion trajectories for an additional test patient in the zero-mean pose frame. Rotations use the rotation-vector representation.}
\label{fig:5}
\end{figure}

Mean RPE decreased from 6.12 to 0.27\,mm during inference (Fig.~\ref{fig:4}(b)), compared with 0.84\,mm for JRM-ADM and 2.26\,mm for learned autofocus (Table~\ref{tab:1}). The reduction relative to autofocus occurred in all 30 patients. Table~\ref{tab:2} reports cohort errors in the individual pose components, and Fig.~\ref{fig:5} shows trajectories for an additional test patient.

\begin{table}[tbp]
\caption{Per-component motion MAE over 30 patients in the zero-mean pose frame (translations: mm; rotation vectors: degrees). Bold indicates column minima.}
\label{tab:2}
\centering
\normalsize
\setlength{\tabcolsep}{8pt}
\begin{tabular}{@{}lrrrrrr@{}}
\toprule
Method & $t_x$ & $t_y$ & $t_z$ & $r_x$ & $r_y$ & $r_z$ \\
\midrule
Uncorrected & 1.93 & 2.08 & 1.95 & 2.01 & 1.90 & 2.17 \\
\shortstack[l]{Learned autofocus\\\cite{ref3}} & 1.30 & 1.33 & 0.36 & 0.95 & 0.98 & 1.09 \\
JRM-ADM \cite{ref7} & 2.17 & 1.97 & 0.06 & 0.09 & 0.07 & 0.68 \\
\textbf{Proposed} & \textbf{0.62} & \textbf{0.61} & \textbf{0.02} & \textbf{0.03} & \textbf{0.03} & \textbf{0.26} \\
\bottomrule
\end{tabular}
\end{table}

The proposed method achieved the lowest cohort MAE in all six components, with translation MAE of at most 0.62\,mm and rotation MAE of at most 0.26\textdegree{}. Its largest residual errors were in the in-plane translations and gantry-axis rotation. JRM-ADM recovered axial translation and out-of-plane rotations with low error, but its in-plane translation MAEs were 2.17 and 1.97\,mm despite an RPE of 0.84\,mm. Thus, low detector-plane RPE did not imply uniformly low component errors; the two measures provide complementary assessments of motion recovery.

\subsection{Ablation study}\label{c.-ablation-study}

\paragraph{Training bridge.}\label{training-bridge.-1}

\begin{figure}[tbp]
\centering
\includegraphics[width=0.58\linewidth]{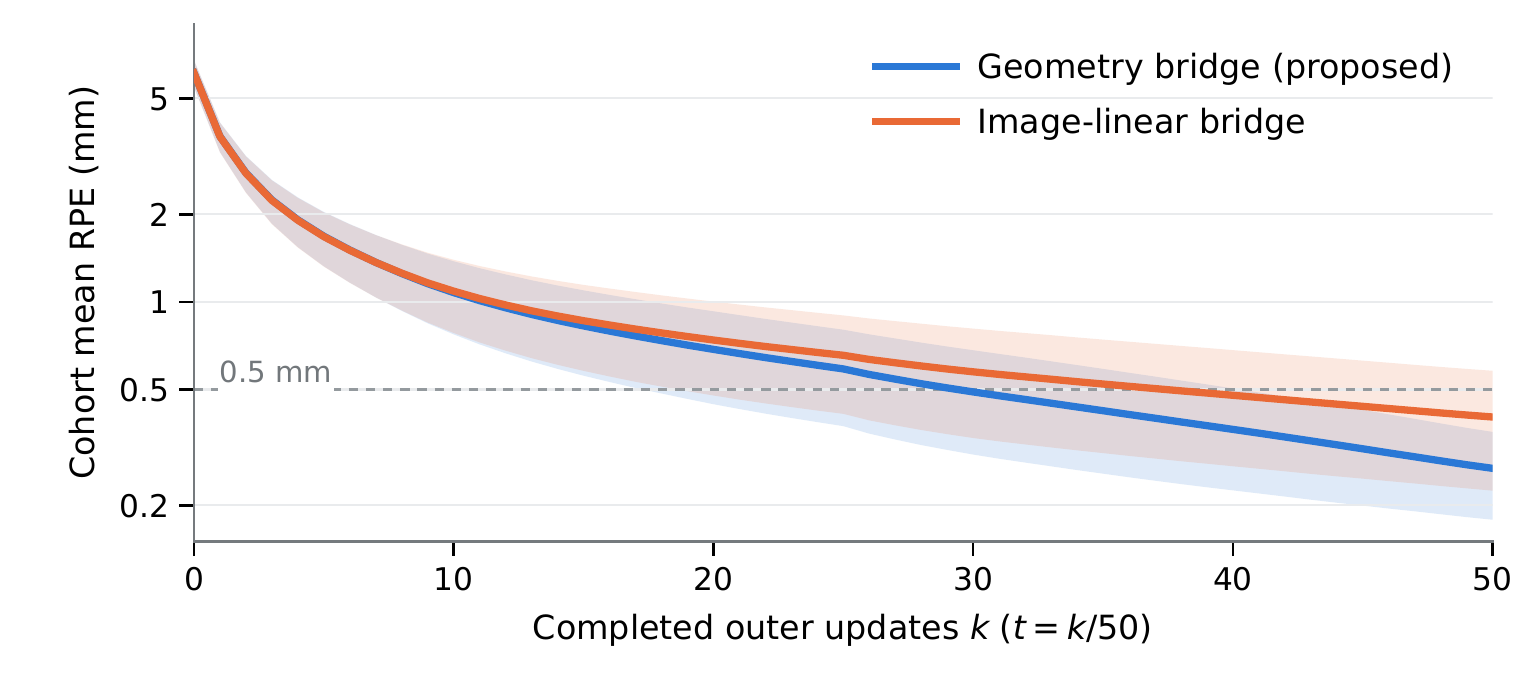}
\caption{Mean RPE in the training-bridge ablation (30 patients). Shading denotes \(\pm\) one standard deviation; update zero is the zero-motion initialization.}
\label{fig:6}
\end{figure}

\changed{With the network and inference settings matched, Table~\ref{tab:3} reports final outcomes and Fig.~\ref{fig:6} shows how motion recovery progresses.}

\begin{table}[tbp]
\caption{Training-bridge ablation with fixed Akima + GD. Final-iterate metrics over 30 patients (mean \(\pm\) standard deviation); bold indicates best means.}
\label{tab:3}
\centering
\normalsize
\setlength{\tabcolsep}{8pt}
\begin{tabular}{@{}lccc@{}}
\toprule
Training bridge & PSNR (dB) & SSIM & RPE (mm) \\
\midrule
Geometry & \textbf{37.12} \(\pm\) \textbf{1.65} & \textbf{0.980} \(\pm\) \textbf{0.010} & \textbf{0.268} \(\pm\) \textbf{0.090} \\
Image-linear & 35.53 \(\pm\) 2.24 & 0.968 \(\pm\) 0.016 & 0.402 \(\pm\) 0.178 \\
\bottomrule
\end{tabular}
\end{table}

The geometry bridge achieved 1.6\,dB higher PSNR and 33\% lower mean RPE. Each of PSNR, SSIM and RPE favored it in 28 of the 30 patient pairs. Under the same 50-update loop, cohort-mean RPE first reached 0.5\,mm or below after 30 updates with the geometry bridge and 38 with the image-linear bridge.

\paragraph{Motion-estimation scheme.}\label{motion-estimation-scheme.-1}

We next replace Akima + GD with B-spline + RMSprop while keeping the geometry-bridge prior fixed (Table~\ref{tab:4}).

\begin{table}[tbp]
\caption{Pose-fitting ablation with the geometry-bridge prior fixed. Final-iterate metrics over 30 patients (mean \(\pm\) standard deviation); bold indicates best means.}
\label{tab:4}
\centering
\normalsize
\setlength{\tabcolsep}{8pt}
\begin{tabular}{@{}lccc@{}}
\toprule
\shortstack[l]{Pose-fitting\\scheme} & PSNR (dB) & SSIM & RPE (mm) \\
\midrule
Akima + GD & \textbf{37.12} \(\pm\) \textbf{1.65} & \textbf{0.980} \(\pm\) \textbf{0.010} & \textbf{0.268} \(\pm\) \textbf{0.090} \\
\shortstack[l]{B-spline +\\RMSprop} & 36.93 \(\pm\) 1.55 & 0.978 \(\pm\) 0.009 & 0.270 \(\pm\) 0.079 \\
\bottomrule
\end{tabular}
\end{table}

Performance remained similar across the two tested pose-fitting schemes. Both retained higher mean PSNR and SSIM and lower mean RPE than learned autofocus and JRM-ADM (Tables~\ref{tab:4} and~\ref{tab:1}).

\subsection{Computational cost}\label{d.-computational-cost}

The proposed method averaged 9.1 minutes per patient, compared with 1.9 minutes for learned autofocus and 4.6 hours for JRM-ADM (Table~\ref{tab:5}).

\begin{table}[tbp]
\caption{Mean total runtime on an RTX A6000 (360 views), including initialization, reconstruction, evaluation and output. Network evaluations count complete predictions, including all patches.}
\label{tab:5}
\centering
\normalsize
\setlength{\tabcolsep}{8pt}
\begin{tabular}{@{}lcc@{}}
\toprule
Method & \shortstack{Network\\evaluations} & \shortstack{Time per\\patient} \\
\midrule
Learned autofocus \cite{ref3} & 100 & 1.9\,min \\
JRM-ADM \cite{ref7} & 100 & 4.6\,h \\
Proposed & 50 & 9.1\,min \\
\bottomrule
\end{tabular}
\end{table}

These timings compare complete implementations; network evaluation counts alone do not capture differences in architecture and projection workload.

\section{Discussion}\label{v.-discussion}

The agreement between lower pose errors and improved FDK reconstructions indicates that the gains extend beyond learned changes in image appearance (Tables~\ref{tab:1} and~\ref{tab:2}). The further improvement from estimated-pose FDK to the final iterate shows the contribution of image refinement after motion compensation. Together, these results support the cooperative role of the learned prior and projection-based updates in joint image and pose recovery.

The displayed JRM-ADM reconstructions retain prominent bone boundaries but show reduced low-contrast soft-tissue detail (Fig.~\ref{fig:3}). The original sparse-view study also reported low-contrast inconsistencies and suggested oversmoothing by the denoising network as a possible explanation \cite{ref7}. Reduced contrast remains visible under our 360-view protocol, despite denser angular sampling than the original 20- and 60-view settings. Our controlled experiments confirmed that the denoised prediction favors soft-tissue smoothing and that the penalty toward this prediction competes with data-fidelity-driven contrast recovery. The geometry bridge trains image updates on changes induced by reducing motion, and its combination with projection-based pose fitting and image refinement yielded better image and motion recovery under the evaluated protocol.

Under the common motion conditions defined in Sec.~\ref{a.-data-and-motion-simulation}, learned autofocus reduced artifacts but retained streaks and missed motion excursions (Figs.~\ref{fig:3} and~\ref{fig:5}). Its learned image-quality score must guide pose updates despite substantial motion corruption of the reconstruction. The residual artifacts and trajectory errors highlight the limitations of this quality-guided optimization under the tested conditions.

The matched bridge ablation shows that training endpoints alone do not determine the prior's effectiveness. With the same network and inference settings, learning intermediate states and velocities induced by decreasing acquisition motion improved both image quality and pose accuracy (Table~\ref{tab:3}). The earlier reduction of mean RPE below 0.5\,mm further shows that this training path supports more accurate recovery within the same update budget (Fig.~\ref{fig:6}). The gains also persisted with both tested pose-fitting schemes (Table~\ref{tab:4}), supporting the use of the geometry-bridge prior with either estimator.

Runtime depends on both network evaluation and repeated projection operations. The proposed loop combines 50 prior evaluations with short CG updates and motion fitting on sampled views at coarse-to-fine resolution, limiting the work per iteration. Its higher image and motion accuracy than JRM-ADM was obtained at approximately one-thirtieth of the runtime (Table~\ref{tab:5}). This is a comparison of complete implementations with different architectures, projectors and solvers, so the speed difference cannot be attributed to the training path alone. Learned autofocus remained faster, with lower image and motion accuracy under the tested protocol.

Limitations include simulated projections, one motion realization per test patient and a single acquisition geometry. Although the source volumes are clinical CT images, the projections do not establish performance on acquired CBCT. The rigid head model excludes independent mandibular and neck motion. Validation with acquired CBCT and broader motion and geometry conditions is needed to assess recovery under clinical noise, scatter, beam hardening and motion outside the training distribution. Soft-tissue observations also require clinical image-quality evaluation.

\section{Conclusion}\label{vi.-conclusion}

Training a 3D flow-matching prior on image changes induced by decreasing acquisition motion improved joint image and pose recovery in head CBCT. On simulated 360-view scans from 30 held-out patients, the proposed framework outperformed the evaluated learned autofocus and JRM-ADM methods in image quality and motion accuracy. The matched bridge ablation and results with two pose-fitting schemes support the contribution of motion-directed training to these gains. Validation on acquired CBCT is the next step toward assessing clinical use.

{\normalsize
\bibliographystyle{ieeenat_fullname}
\bibliography{references}

@article{ref1,
  author = {Ko, Y. and Moon, S. and Baek, J. and Shim, H.},
  title = {{Rigid and non-rigid motion artifact reduction in X-ray CT using attention module}},
  journal = {Med. Image Anal.},
  volume = {67},
  pages = {101883},
  year = {2021}
}

@article{ref2,
  author = {Sisniega, A. and Stayman, J. W. and Yorkston, J. and Siewerdsen, J. H. and Zbijewski, W.},
  title = {{Motion compensation in extremity cone-beam CT using a penalized image sharpness criterion}},
  journal = {Phys. Med. Biol.},
  volume = {62},
  number = {9},
  pages = {3712--3734},
  year = {2017}
}

@article{ref3,
  author = {Thies, M. and others},
  title = {{A gradient-based approach to fast and accurate head motion compensation in cone-beam CT}},
  journal = {IEEE Trans. Med. Imaging},
  volume = {44},
  number = {2},
  pages = {1098--1109},
  year = {2025}
}

@article{ref4,
  author = {Huang, H. and Siewerdsen, J. H. and Zbijewski, W. and Weiss, C. R. and Unberath, M. and Ehtiati, T. and Sisniega, A.},
  title = {{Reference-free learning-based similarity metric for motion compensation in cone-beam CT}},
  journal = {Phys. Med. Biol.},
  volume = {67},
  number = {12},
  pages = {125020},
  year = {2022}
}

@inproceedings{ref5,
  author = {Chung, H. and Kim, J. and McCann, M. T. and Klasky, M. L. and Ye, J. C.},
  title = {{Diffusion posterior sampling for general noisy inverse problems}},
  booktitle = {ICLR},
  year = {2023}
}

@inproceedings{ref6,
  author = {Chung, H. and Lee, S. and Ye, J. C.},
  title = {{Decomposed diffusion sampler for accelerating large-scale inverse problems}},
  booktitle = {ICLR},
  year = {2024}
}

@article{ref7,
  author = {De Paepe, A. and Bousse, A. and Phung-Ngoc, C. and Mellak, Y. and Visvikis, D.},
  title = {{Adaptive diffusion models for sparse-view motion-corrected head cone-beam CT}},
  journal = {IEEE Trans. Radiat. Plasma Med. Sci.},
  volume = {10},
  number = {5},
  pages = {662--672},
  year = {2026}
}

@inproceedings{ref8,
  author = {Friedrich, P. and Wolleb, J. and Bieder, F. and Durrer, A. and Cattin, P. C.},
  title = {{WDM: 3D wavelet diffusion models for high-resolution medical image synthesis}},
  booktitle = {MICCAI Workshop on Deep Generative Models},
  pages = {11--21},
  year = {2024}
}

@inproceedings{ref9,
  author = {Lipman, Y. and Chen, R. T. Q. and Ben-Hamu, H. and Nickel, M. and Le, M.},
  title = {{Flow matching for generative modeling}},
  booktitle = {ICLR},
  year = {2023}
}

@article{ref10,
  author = {Feldkamp, L. A. and Davis, L. C. and Kress, J. W.},
  title = {{Practical cone-beam algorithm}},
  journal = {J. Opt. Soc. Am. A},
  volume = {1},
  number = {6},
  pages = {612--619},
  year = {1984}
}

@inproceedings{ref11,
  author = {Venkatakrishnan, S. V. and Bouman, C. A. and Wohlberg, B.},
  title = {{Plug-and-play priors for model based reconstruction}},
  booktitle = {IEEE GlobalSIP},
  pages = {945--948},
  year = {2013}
}

@article{ref12,
  author = {Preuhs, A. and Manhart, M. and Roser, P. and Hoppe, E. and Huang, Y. and Psychogios, M. and Kowarschik, M. and Maier, A.},
  title = {{Appearance learning for image-based motion estimation in tomography}},
  journal = {IEEE Trans. Med. Imaging},
  volume = {39},
  number = {11},
  pages = {3667--3678},
  year = {2020}
}

@article{ref13,
  author = {Akima, H.},
  title = {{A new method of interpolation and smooth curve fitting based on local procedures}},
  journal = {J. ACM},
  volume = {17},
  number = {4},
  pages = {589--602},
  year = {1970}
}

@article{ref14,
  author = {Chilamkurthy, S. and others},
  title = {{Deep learning algorithms for detection of critical findings in head CT scans: a retrospective study}},
  journal = {The Lancet},
  volume = {392},
  number = {10162},
  pages = {2388--2396},
  year = {2018}
}

@misc{ref15,
  author = {Yang, T. and Hu, J. and Fessler, J. A. and Shen, L.},
  title = {{Local patches meet global context: scalable 3D diffusion priors for computed tomography reconstruction}},
  howpublished = {arXiv:2512.18161},
  year = {2025}
}

@inproceedings{ref16,
  author = {Kim, H. and Champley, K.},
  title = {{Differentiable forward projector for X-ray computed tomography}},
  booktitle = {ICML Workshop on Differentiable Almost Everything},
  year = {2023},
  note = {arXiv:2307.05801; LEAP, \url{https://github.com/LLNL/LEAP}}
}

@article{ref17,
  author = {Wang, Z. and Bovik, A. C. and Sheikh, H. R. and Simoncelli, E. P.},
  title = {{Image quality assessment: From error visibility to structural similarity}},
  journal = {IEEE Trans. Image Process.},
  volume = {13},
  number = {4},
  pages = {600--612},
  year = {2004}
}

@article{hyun2026diffusioninr,
  author = {Hyun, Subong and Yun, Sungho and Lee, Seoyoung and Choi, Da-in and Cho, Seungryong},
  title = {{Diffusion prior-guided implicit neural representation for metal artifact reduction in sparse-view CT reconstruction}},
  journal = {Physics in Medicine and Biology},
  volume = {71},
  number = {10},
  pages = {105007},
  year = {2026},
  doi = {10.1088/1361-6560/ae6a5a}
}

@inproceedings{chung2023diffusionmbir,
  author = {Chung, Hyungjin and Ryu, Dohoon and McCann, Michael T. and Klasky, Marc L. and Ye, Jong Chul},
  title = {{Solving 3D Inverse Problems Using Pre-Trained 2D Diffusion Models}},
  booktitle = {Proceedings of the IEEE/CVF Conference on Computer Vision and Pattern Recognition (CVPR)},
  year = {2023},
  pages = {22542--22551}
}

@inproceedings{martin2025pnpflow,
  author = {Martin, S{\'e}gol{\`e}ne and Gagneux, Anne and Hagemann, Paul and Steidl, Gabriele},
  title = {{PnP-Flow: Plug-and-Play Image Restoration with Flow Matching}},
  booktitle = {ICLR},
  year = {2025}
}

@inproceedings{cicek2016unet3d,
  author = {{\c{C}}i{\c{c}}ek, {\"O}zg{\"u}n and Abdulkadir, Ahmed and Lienkamp, Soeren S. and Brox, Thomas and Ronneberger, Olaf},
  title = {{3D U-Net: Learning Dense Volumetric Segmentation from Sparse Annotation}},
  booktitle = {Medical Image Computing and Computer-Assisted Intervention -- MICCAI 2016},
  series = {Lecture Notes in Computer Science},
  volume = {9901},
  pages = {424--432},
  publisher = {Springer},
  year = {2016},
  doi = {10.1007/978-3-319-46723-8_49}
}

@inproceedings{loshchilov2019adamw,
  author = {Loshchilov, Ilya and Hutter, Frank},
  title = {{Decoupled Weight Decay Regularization}},
  booktitle = {International Conference on Learning Representations (ICLR)},
  year = {2019},
  url = {https://openreview.net/forum?id=Bkg6RiCqY7}
}

@inproceedings{liu2023rectifiedflow,
  author = {Liu, Xingchao and Gong, Chengyue and Liu, Qiang},
  title = {{Flow Straight and Fast: Learning to Generate and Transfer Data with Rectified Flow}},
  booktitle = {International Conference on Learning Representations (ICLR)},
  year = {2023},
  url = {https://openreview.net/forum?id=XVjTT1nw5z}
}

@inproceedings{kawar2022ddrm,
  author = {Kawar, Bahjat and Elad, Michael and Ermon, Stefano and Song, Jiaming},
  title = {{Denoising Diffusion Restoration Models}},
  booktitle = {Advances in Neural Information Processing Systems},
  volume = {35},
  pages = {23593--23606},
  year = {2022}
}

@inproceedings{zhang2017ircnn,
  author = {Zhang, Kai and Zuo, Wangmeng and Gu, Shuhang and Zhang, Lei},
  title = {{Learning Deep CNN Denoiser Prior for Image Restoration}},
  booktitle = {Proceedings of the IEEE Conference on Computer Vision and Pattern Recognition (CVPR)},
  pages = {3929--3938},
  year = {2017}
}

@inproceedings{zhu2023diffpir,
  author = {Zhu, Yuanzhi and Zhang, Kai and Liang, Jingyun and Cao, Jiezhang and Wen, Bihan and Timofte, Radu and Van Gool, Luc},
  title = {{Denoising Diffusion Models for Plug-and-Play Image Restoration}},
  booktitle = {Proceedings of the IEEE/CVF Conference on Computer Vision and Pattern Recognition (CVPR) Workshops},
  pages = {1219--1229},
  year = {2023}
}

@inproceedings{wang2023ddnm,
  author = {Wang, Yinhuai and Yu, Jiwen and Zhang, Jian},
  title = {{Zero-Shot Image Restoration Using Denoising Diffusion Null-Space Model}},
  booktitle = {The Eleventh International Conference on Learning Representations},
  year = {2023}
}

@inproceedings{ho2020ddpm,
  author = {Ho, Jonathan and Jain, Ajay N. and Abbeel, Pieter},
  title = {{Denoising Diffusion Probabilistic Models}},
  booktitle = {Advances in Neural Information Processing Systems},
  volume = {33},
  pages = {6840--6851},
  year = {2020}
}
}

\end{document}